\documentstyle[12pt,a41,epsfig]{article}

\newcommand\GeV{\,\mbox{GeV}}

\begin{document}
\setlength{\baselineskip}{0.515cm}
\sloppy
\thispagestyle{empty}
\begin{flushleft}
DESY 98--170 \hfill
{\tt hep-ph/9811519}\\
November 1998\\
\end{flushleft}

\setcounter{page}{0}

\mbox{}
\vspace*{\fill}
\begin{center}{
\LARGE\bf Less Singular Terms and Small \boldmath{$x$} Evolution}

\vspace{2mm}
{\LARGE\bf in a Soluble Model}$^{
^{\footnotemark}}$\footnotetext{Work supported in part by EU contract 
FMRX-CT98-0194(DG 12 - MIHT)}

\vspace{5em}
\large
J. Bl\"umlein~~~and~~~W. L. van Neerven\footnote{On leave of
absence from Instituut-Lorentz, University of Leiden, P.O. Box 9506
, 2300 RA Leiden, The Netherlands}
\\
\vspace{5em}
\normalsize
{\it DESY--Zeuthen}\\
{\it Platanenallee 6, D--15735 Zeuthen, Germany}\\
\today
\vspace*{\fill}
\end{center}
\begin{abstract}
\noindent
We calculate the effect of the less singular terms at small $x$ on the 
evolution of the coefficient function in $\phi^3$ theory in six
dimensions, which result from a complete solution of the ladder equation.
Scale-invariant next-to-leading order contributions are also studied.
We show that the small--$x$ approximation does not deliver the dominant
contributions.
\end{abstract}

\vspace{1mm}
\noindent
\begin{center}
PACS numbers~:~12.38.Cy

\end{center}

\vspace*{\fill}
\newpage

\vspace{1mm}
\noindent
The resummation of the terms of the type $(\ln^i x)/x$~\cite{BFKL1,BFKL2},
occurring in the flavor singlet evolution kernels of the deep inelastic 
scattering structure functions at small values of $x = Q^2/(2p.q)$,
leads to large effects~\cite{BV1,BV2}. If one includes the next-to-leading
order small $x$ terms, the effects are so strong that the resummed gluon
splitting function becomes already negative at $x$--values as large as
$10^{-2}$ for $Q^2 \simeq 20 (\GeV/c)^2$~\cite{BV2}. Moreover the 
corrections
to the coefficient $\omega$ describing the large energy ($\sqrt s$) 
behavior of the inclusive cross sections, given by $s^{\omega}$, may 
lead to negative values of $\omega$ \cite{BRN1}. Already 
in~\cite{SUBL1} the question was raised to which extent the leading
logarithms dominate the perturbation series of the kernels or that 
less--singular terms are also important.
For the case of the splitting functions and coefficient functions, out of 
which the kernels are constituted, this is shown in fixed--order 
perturbation 
theory in \cite{BV2,SUBL1}. A relation  between the most singular and 
less
singular terms at $x=0$ is for instance given by the conservation
laws, as fermion number~\cite{BV3} and energy--momentum 
conservation~\cite{BV2,EHW}. As outlined in Refs.~\cite{BV2,BNRV} the 
less 
singular contributions may cure the large effects in the evolution
 kernels
caused by the resummation of the leading logarithms mentioned above.

In QCD the resummation of the most singular term in the leading order 
kernel 
is achieved by the leading order BFKL-equation~\cite{BFKL1}. 
Unfortunately
until now one is not able to resum the full kernel so that the subleading
contributions cannot be included in the resummation. Therefore one has no real
check on the validity of the small $x$ approximation for the high energy
behavior of the cross section or the small $x$ behavior of the structure
functions. However in the case of massless $\phi_6^3$-theory the 
resummation of the exact leading order kernel is known \cite{LOVEL} so 
that the validity of the small-$x$ approximation can be tested here. 
This will be our goal in this paper where we obtain information about the
r\^ole of the less-singular terms in the kernel which are neglected in 
the BFKL approach.
In particular we study the difference between the exact and approximate
resummation for the splitting functions and the coefficient functions
where we also discuss some contributions from the next-to-leading order
kernel.

The Bethe--Salpeter equation for the forward scattering amplitude 
$T(p,q)$ in massless $\phi^3$-theory is given by
\begin{eqnarray}
T(q,p) = \frac{2^{2-D}}{\pi^{D/2} \Gamma\left[(D-2)/2\right]}
\frac{\lambda^2}{(q-p)^2} + \frac{1}{(2\pi)^D} \int d^D k
\frac{\lambda^2 T(k,p)}{(q-k)^2\left[k^2\right]^2}~.
\label{eqBSE}
\end{eqnarray}
Here, $\lambda$ denotes the coupling constant and $D$ is the dimension 
of space--time. Furthermore the leading order kernel is given by the
propagator $1/(q-p)^2$. For $D~=~6$ the quantity $q.p~T(q,p)$ becomes 
scale invariant. The equation above is solved by expanding all functions 
in terms of Gegenbauer polynomials which is the analogue of the partial 
wave expansion in three dimensions in terms of Legendre functions. The 
leading $q^2$-behavior  of the Nth partial wave $T_N(q,p)$ is given 
by~\cite{LOVEL}
\begin{eqnarray}
p.q~T_N(q,p)~&\sim&~p.q~T_N^{(0)}(p,q)~\hat{T}_N\left(\frac{q^2}{p^2}
\right)~\quad p^2<0, \quad q^2<0, \quad \mbox{with}
\nonumber\\[2ex]
p.q~T_N^{(0)}(p,q) &=& \left(\frac{q^2}{p^2}\right)^{-(N+1)/2}, \qquad
\hat{T}_N\left(\frac{q^2}{p^2}\right )
=~\left(\frac{q^2}{p^2}\right)^{-{\gamma}_L(N,a_s)/2} \,,
\label{eqTN}
\end{eqnarray}
with
\begin{eqnarray}
{\gamma}_L(N,a_s) = \sqrt{(N+2)^2+1-2\sqrt{(N+2)^2+4 a_s}}
- (N+1)~= \sum_{k=0}^{\infty} a_s^k \gamma_L^{(k)}(N),
\label{eqAN1}
\end{eqnarray}
and $a_s = \lambda^2/(4\pi)^3 = \alpha_s/(4\pi)$.
The quantity above is nothing but than the ladder approximation to
the anomalous dimension of the composite operator of spin $N$ given by
$\phi \partial_{\mu_1} \cdots  \partial_{\mu_N} \phi$ in 
$\phi_6^3$-theory. This we have verified by a fixed--order calculation
up to three-loop order. 
Because of scale invariance the quantity $\hat{T}_N(q,p)$ satisfies the
scaling equation
\begin{eqnarray}
\left[p \frac{\partial}{\partial p} - \gamma_L(N,a_s)\right]
\hat{T}_N\left(\frac{q^2}{p^2}\right) = 0,
\label{eqREN}
\end{eqnarray}
which equals to the renormalization group equation in the conformal 
limit. 
The splitting functions and the coefficient functions, both depending on
the variable $x$, are related to the quantities above via a Mellin 
transformation. The splitting function is given by
\begin{equation}
\label{eqMEL}
\gamma(N,a_s) = - \int_0^1 dx x^{N-1} P(x,a_s)
\end{equation}
and the $Q^2$-dependence of the coefficient function $C$ is determined
by the anomalous dimension via the relation
\begin{equation}
\label{eqCO}
\left(\frac{Q^2}{Q_0^2}\right)^{-\gamma(N,a_s)/2} =
\int_0^1 dx x^{N-1} C(x,Q^2,\alpha_s), \qquad q^2=-Q^2 \,,
\end{equation}
which satisfies Eq. (\ref{eqREN}).

Since the full leading order kernel has been resummed by Eq.~(\ref{eqBSE})
the lowest order coefficient of the ladder approximation 
coincides, up to the constant $1/6$ due to energy--momentum conservation,
with the exact one given by $\gamma_{SS}^{(0)}(N)$,
\begin{eqnarray}
\label{eqAN0}
\gamma_{SS}^{(0)}(N) =  - \frac{2}{(N+1)(N+2)} + \frac{1}{6}~.
\end{eqnarray}
However, no resummation exists for the next-to-leading kernel and we only
know the exact order $\alpha_s^2$-contribution to the anomalous 
dimension~\cite{KBFK} which are given in the $\overline{\rm MS}$-scheme by
\begin{eqnarray}
\label{eqGA1}
\gamma_{SS}^{(1)}(N) &=& \gamma_{SS}^{(1,a)}(N)
                      +  \frac{1}{2}\left[1+(-1)^N\right]
\gamma_{SS}^{(1,b)}(N)
\end{eqnarray}
with
\begin{eqnarray}
\label{GAa}
\gamma_{SS}^{(1,a)}(N) &=&
\frac{5}{3}~\frac{S_1(N)}{(N+1)(N+2)}
- \frac{1}{6}~\frac{22+111 N +211 N^2 + 138 N^3 + 28 N^4}
{(N+1)^3 (N+2)^3}
+ \frac{13}{216}~,
\\
\label{GAb}
\gamma_{SS}^{(1,b)}(N) &=&
- \frac{2}{(N+1)^2(N+2)^2}~.
\end{eqnarray}
Here $\gamma_{SS}^{(1,a)}(N)$ is due to the ladder-, vertex-, and
self energy contributions. The contribution $\gamma_{SS}^{(1,b)}(N)$ is 
obtained from the crossed--ladder graph which like the planar ladder 
graphs, leading to $\gamma_L(N,a_s)$, satisfies the scaling equation
(\ref{eqREN}). From the Mellin transforms in 
Eqs.~(\ref{eqMEL},\ref{eqCO}) and the anomalous dimensions given above 
one infers that the most singular behavior of the splitting functions and
the coefficient functions at $x=0$ is determined in $\phi_6^3$--theory by
the leading pole terms of the type $O\left[(a_s/(N+1))^k\right]$. Notice 
that the latter are shifted by two units with respect to those given in 
QCD of $O\left[(a_s/(N-1))^k\right]$. Taking the limit
$N \rightarrow -1$ in $\gamma_{SS}^{(i)}(N,a_s)$ ($i=0,1$) one finds that
the most singular contributions $\propto [a_s/(N+1)]^k$ are all contained
in the ladder approximation $\gamma_L(N,a_s)$.

In analogy to the leading order BFKL-equation~\cite{BFKL1} one may 
determine now the all-order small-$x$ limit of the coefficient function
via Eq.~(\ref{eqCO}). For this purpose we expand $\gamma_L(N,a_s)$ into 
an infinite series in $a_s$ and collect all terms of 
$O\left[(a_s/(N+1))^k\right]$, which can be represented in a closed form 
again~\cite{SLPL},
\begin{eqnarray}
{\gamma}_{L,N \rightarrow -1}(N,a_s) =
\sum_{k=1}^{\infty} c_{0,k} \frac{a_s^k}{(N+1)^{2k-1}}
\equiv (N+1) \left [\sqrt{1 - \frac{4 a_s}{(N+1)^2}} - 1 \right]~.
\label{eqAN2}
\end{eqnarray}
Performing the same series expansion on the exact ladder approximation
in~Eq.~(\ref{eqAN1}) one can study the size of all terms which are less
singular at $N=1$ than those contained in the approximate expression
above. The series expansion of Eq.~(\ref{eqAN1}) is given by
\begin{equation}
\label{eqGALe}
\gamma_L(N,a_s) = \sum_{k=1}^{\infty} \sum_{l=0}^{\infty}
c_{l,k} \frac{a_s^k}{(N+1)^{2k-1+l}}~.
\end{equation}
A comparison between the residues of the leading pole terms given by
$c_{0,k}$ and the residues corresponding to the subleading contributions
$c_{l,k}$ ($l \ge 1$) is made in table 1.
\begin{table}\centering
\begin{tabular}{||c||r|r|r|r|r||}
\hline\hline
\multicolumn{1}{||c||}{$k$}&
\multicolumn{1}{c|}{$c_{0,k}$}&
\multicolumn{1}{c|}{$c_{1,k}$}&
\multicolumn{1}{c|}{$c_{2,k}$}&
\multicolumn{1}{c|}{$c_{3,k}$}&
\multicolumn{1}{c||}{$c_{4,k}$}\\
\hline\hline
 1&        $-$2&      2&      $-$2&        2&       $-$2\\
 2&        $-$2&      4&      $-$4&        2&          2\\
 3&        $-$4&     12&     $-$20&       24&      $-$24\\
 4&       $-10$&     40&     $-$88&      140&     $-$180\\
 5&       $-$28&    140&    $-$380&      740&    $-$1156\\
 6&       $-$84&    504&   $-$1624&     3724&    $-$6804\\
 7&      $-$264&   1848&   $-$6888&    18144&   $-$37856\\
 8&      $-$858&   6864&  $-$29040&    86328&  $-$202248\\
 9&     $-$2860&  25740& $-$121863&   403260& $-$1047420\\
10&     $-$9724&  97240& $-$509080&  1856140& $-$5291572\\
\hline \hline 
\end{tabular}

\vspace{4mm}
\noindent
{\sf
table~1~: The first expansion coefficients of Eq.~(\ref{eqGALe}).}
\end{table}
From the table we infer that the residues of the less singular pole 
terms 
show alternating signs and are of larger modulus. Therefore the leading 
series in Eq. (\ref{eqAN2}) receives large corrections and is not
dominant. A similar behavior was 
observed for the anomalous dimensions and coefficient functions in 
fixed--order perturbation theory in QCD before~\cite{BV2,SUBL1}.

To illustrate the importance of the subleading pole terms we have 
computed various splitting functions from the anomalous dimensions 
listed above by inverse Mellin transformation (see Eq. (\ref{eqMEL}))
for $ 10^{-10} < x < 1$ in figure~1 for $\alpha_s = 0.2$. All curves are
normalized to the leading order splitting function for $x<1$, 
$a_s P_L^{(0)}(x) = 2 a_s x(1-x)$. The splitting function in
next-to-leading order $a_sP_{SS}^{(0)}(x)+a_s^2P_{SS}^{(1)}(x)$
(Eqs. (\ref{eqAN0},\ref{eqGA1})) shows
a relative rise towards small $x$ compared
to the leading order result Eq.~(\ref{eqGA1}).
The resummation of all ladder graphs, $P_L$~Eq.(\ref{eqAN1}),
reveals an appreciable contribution
to the exact order $\alpha_s^2$ corrected splitting function. However
this effect is not due to the leading pole terms 
$O\left[(a_s/(N+1))^k\right]$ or most singular $x$-contributions 
$O(x \ln^{k-1} x)$. This is shown by the plot of the small-$x$
approximation ${ P}_{L, x \rightarrow 0}$ which is derived from
Eq.~(\ref{eqAN2}). Here we see a large discrepancy between the exact 
ladder resummation (Eq. (\ref{eqAN1})) and its small-$x$ or leading pole 
approximation (Eq. (\ref{eqAN2})). This becomes even larger when $x$ gets
smaller, where the approximation above is supposed to hold.

The features shown for the splitting functions also hold for the 
coefficient functions. This is revealed in figure~2 where we have plotted
the ratio $C_{L,x \rightarrow 0}(x,Q^2,\alpha_s)/C_L(x,Q^2,\alpha_s)$. 
Here $C_L$ and $C_{L,x \rightarrow 0}$ are determined via Eq.~(\ref{eqCO}) 
by Eq. (\ref{eqAN1}) and Eq. (\ref{eqAN2}), respectively. The ratio
illustrates the behavior of the structure function idealized by a 
$\delta(1-x)$--like parton distribution in the conformal limit.
Like in the case of the splitting functions the discrepancy between the 
exact and small-$x$ resummation becomes larger when $x$ gets smaller. 
This reveals again that the leading pole terms or small-$x$ terms do not 
constitute the bulk of the resummed ladder-solution. The next-to-leading
order solution of the Bethe-Salpeter equation in (\ref{eqBSE}) also 
involves the resummation of the crossed ladder graph, of which the order
$\alpha_s^2$ contribution to the anomalous dimension is given in
Eq.~(\ref{GAb}). Addition of the latter contribution
to $\gamma_L$ in 
Eq.~(\ref{eqAN1}) leads to a positive contribution to $C_L$ so that the
ratio decreases. The latter also happens when $Q^2$ in Eq.~(\ref{eqCO}) 
or the fixed coupling constant $\alpha_s$ gets smaller. 

In summary it was shown that in $\phi^3_6$-theory the less singular
contributions in the anomalous dimension, which are obtained by the
complete ladder solution of the Bethe-Salpeter equation
lead to sizeable corrections to the
resummation of the small-$x$ terms in Eq.~(\ref{eqAN2}). These
contributions lead to a smaller rise of the coefficient function
at small $x$. One might wish to obtain a similar complete conformal 
solution also in QCD. As signaled by the behavior of the
fixed--order results such a resummation could lead to a more stable
solution at small $x$ than shown by the currently 
available resummations~\cite{BFKL1,BFKL2}.

\vspace{1mm}
\noindent
{\bf Acknowledgment.}~~For  discussions we would like to thank
V. Ravindran and A. Vogt.



\newpage
\begin{center}

\mbox{\epsfig{file=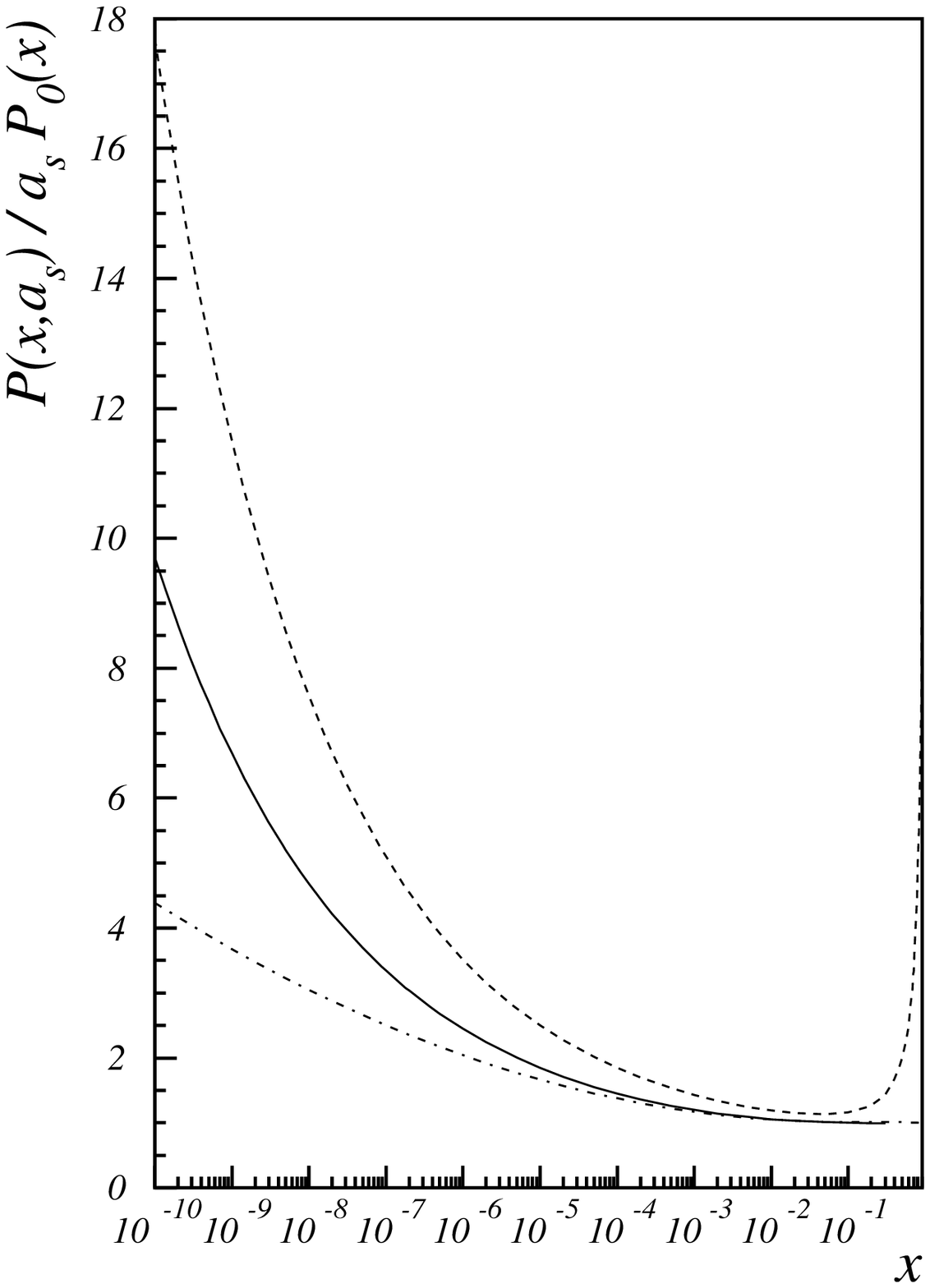,height=18cm,width=16cm}}

\vspace{2mm}
\noindent
\small
\end{center}
{\sf
Figure~1:~Fixed--order and resummed splitting functions 
$P(x,a_s)$ normalized to $a_s P_{SS}^{(0)}(x)$ for $\alpha_s = 0.2$. 
Dash-dotted line : $P=a_sP_L^{(0)}+a_s^2P_{SS}^{(1)}$
Eqs.~(\ref{eqAN0}),
(\ref{eqGA1}).
Solid line : $P=P_L$, Eq. (\ref{eqAN1}).
Dashed line : $P=P_{L, x \rightarrow 0}$, Eq. (\ref{eqAN2}).
\normalsize
\newpage
\begin{center}

\mbox{\epsfig{file=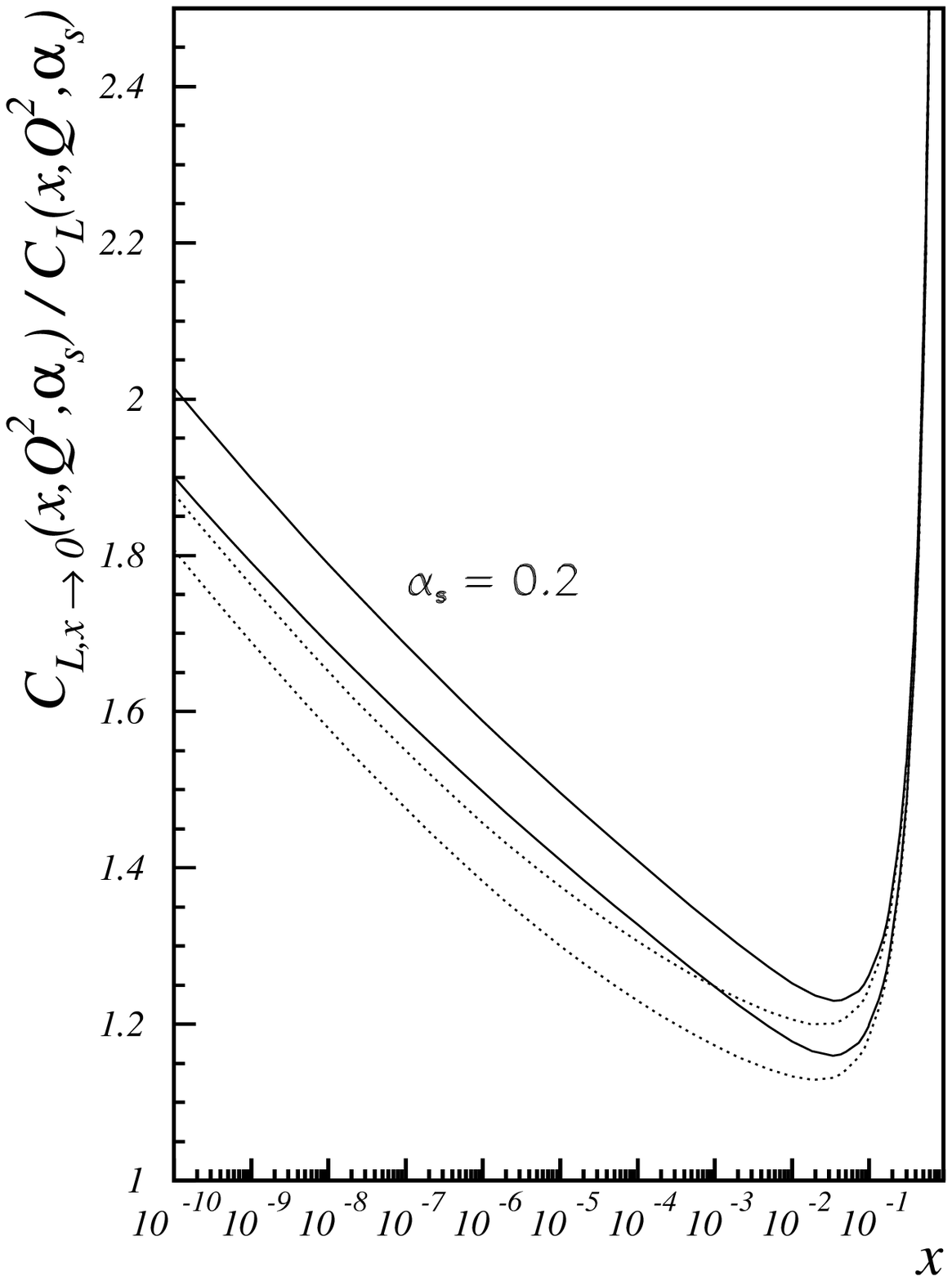,height=18cm,width=16cm}}

\vspace{2mm}
\noindent
\small
\end{center}
{\sf
Figure~2:~Ratio of the  coefficient functions 
$C_{L,x \rightarrow 0}(x,Q^2,\alpha_s)/C_L(x,Q^2,\alpha_s)$, with 
$Q^2_0 = 4 (\GeV/c)^2$ and $\alpha_s = 0.2$. Solid lines~: $C_L$
Eq.~(\ref{eqAN1}), $C_{L,x \rightarrow 0}$ Eq. (\ref{eqAN2}).
Dotted lines : To $C_L$ is added the contribution due to Eq.~(\ref{GAb}). The upper
and lower lines correspond to $Q^2 = 10^4 (\GeV/c)^2$ and $Q^2 = 10^2 
(\GeV/c)^2$ respectively. 
\normalsize
\end{document}